\documentclass{aa}
\usepackage{graphicx}
\begin{document}


   \title{XMM-Newton Observation of a Distant X-ray
Selected Cluster of Galaxies at z=1.26 with Possible Cluster Interaction} 

   \author{Yasuhiro Hashimoto\inst{1}
          G\"unther Hasinger\inst{1}
          Monique Arnaud\inst{2}
          Piero Rosati\inst{3}
          \and
          Takamitsu Miyaji\inst{4}
          }

    \institute{Astrophysikalisches Institut Potsdam, An der Sternwarte 16, 
               D-14482 Potsdam, Germany\\
              \email{hashimot@aip.de}
         \and 
             Service d'Astrophysique
             CEA Saclay, 91191 Gif sur Yvette Cedex, France\\
         \and 
             European Southern Observatory, D-85748 Garching, Germany\\
         \and 
             Department of Physics, Carnegie Mellon University, 
             5000 Forbes Ave., Pittsburgh, PA 15213, USA\\
          }

   \offprints{Y.\ Hashimoto}

   \date{Received Apr. 20, 2001; accepted Oct. 30, 2001}

   \authorrunning{Hashimoto et al.}
   \titlerunning{XMM observation of a high-z double-lobed cluster}

\abstract{
We report on the XMM-Newton observation of RXJ1053.7+5735,
one of the most distant (z = 1.26)
X-ray selected clusters of galaxies, which also shows an unusual 
double-lobed X-ray morphology, 
indicative of possible cluster-cluster interaction.
The cluster was discovered during 
our ROSAT deep pointings in the direction of the Lockman Hole. 
The XMM-Newton observations were performed with the European Photon
Imaging Camera (EPIC) during the performance verification phase. 
Total effective exposure time
was $\sim$ 100 ksec. 
The best fit temperature based on a simultaneous fit of spectra from
the all EPIC cameras (pn+MOS) is 4.9 $^{+1.5}_{-0.9}$ keV.
Metallicity is poorly constrained even using the joint fit of
all spectra,
with an upper limit on the iron abundance of 0.62 solar.
Using the best fit model parameters, we derived an
unabsorbed (0.2-10) keV
flux of $f_{0.2-10}$ =
3.0 $\times$ 10$^{-14}$ erg cm $^{-2}$ s$^{-1}$,
corresponding to 
a bolometric luminosity of
$L_{bol}$ =
3.4 $\times$
10$^{44}$ h$_{50}^{-2}$erg s$^{-1}$.
Despite the fact that it was observed at fairly large off-axis angle,
the temperature errors are much smaller compared with
those of  typical
measurements based on ASCA or Beppo-Sax observations of
high-z (z $>$ 0.6) clusters, 
demonstrating the power of the XMM for
determining the X-ray temperature for high-z clusters.
The measured temperature and luminosity show that
one can easily reach the intrinsically X-ray
faint and cool cluster
regime comparable with those of z $\sim$ 0.4 clusters observed by
past satellites.
The new cluster temperature and L$_{bol}$ we have measured for RXJ1053.7+5735
is consistent with a weak/no  evolution of the L$_{bol}$ - Tx
relation out to  z $\sim$ 1.3, which lends support to a
low $\Omega_{m}$ universe,
although more data-points of z $>$ 1 clusters are required for a more definitive statement.
The caution has to be also exercised
in interpreting the result, because of the uncertainty associated with
the dynamical status of this cluster.
}

   \maketitle

\section{Introduction}

Clusters of galaxies can be identified to high redshift and
can be used as tracers of the evolution of structure.
Since the evolution of structure depends on parameters such as
$\Omega_{m}$, $\Lambda$, \& $P(k)$ in hierarchical cosmologies, 
the study of the
evolution of galaxy clustering provides us with an important constraint on
the cosmology (e.g. Press \& Schechter 1974; Peebles et al. 1989; Eke et al.
1996).
X-ray observations 
provide a powerful and unique means of selecting the clusters and
characterizing their properties, as  
X-ray flux is
proportional to the square
of the electron density, and therefore
less affected 
than optical data
by the superposed structures. 
Measurements of the gas temperatures in high redshift clusters of 
galaxies strongly
constrain cosmological models because cluster temperatures are closely
related to cluster masses, 
and the evolution of the cluster mass function
with redshift is quite sensitive
to cosmological parameters.

X-ray cluster surveys based on ROSAT-PSPC
data (e.g. Rosati et al. 1995, 98; Scharf et al. 1997; Burke et al 1997;
Vikhlinin et al. 1998)
detect sizable samples of distant clusters (z $>$ 0.5).
Although the number of high-z clusters has 
significantly increased,  
intracluster medium (ICM) properties 
(such as X-ray temperature: Tx) at high redshift
are still largely unexplored. This is due to the limited
effective area and spatial resolution of the past X-ray missions;
only long ASCA/Beppo-SAX observations, 
with a broad energy response, have
allowed studies of the high-z cluster ICM properties.
Only a few  high-z (z $>$ 0.6) clusters have
directly measured X-ray temperatures
(e.g. Donahue et al. 1999; Della Ceca et al. 2000).
Moreover, these high-z samples are inevitably biased toward
intrinsically X-ray bright, and thus high temperature clusters, 
which complicates the  
investigation of ICM evolution.

Advent of Chandra, with its 
bigger effective area and higher angular resolution than the past satellites, 
made us possible to effectively study the ICM properties of high-z clusters.
Chandra's high angular resolution, in particular,  
greatly helps us to avoid confusion 
with point sources (e.g. Stanford et al. 2001; Borgani et al. 2001; 
Jeltema et al. 2001; Cagnoni et al. 2001). 
Unfortunately, for those with directly measured Tx, the errors on Tx are 
still relatively large. 
Moreover,  Chandra is not particularly sensitive to
low surface brightness emission,  which is sometimes unfavorable 
for the analysis of faint X-ray features often present in high-z clusters.  

The XMM-Newton (XMM) pn and MOS CCD cameras 
have $\sim$
10 times larger
effective area than ASCA GIS/SIS \&  Beppo-SAX MECS 
(in addition to the fact that it has significantly
sharper PSF than ASCA \& Beppo-Sax).
Even compared with Chandra ACIS, the effective area
is more than 5 times larger
which helps to 
increase the accuracy of the Tx determination per given 
exposure time. 
With its high throughput and moderate (compared with Chandra) 
angular resolution, XMM is extremely sensitive to low surface brightness
X-ray emission. 
These properties make 
XMM pn/MOS very suitable instruments 
to investigate the ICM properties of distant clusters.

In this paper, we report on the XMM observation of 
one of the most distant X-ray selected clusters 
(z = 1.26) 
of galaxies, which also shows an unusual double-lobed 
X-ray morphology. This cluster, 
RXJ1053.7+5735, 
was discovered during our 
deep ROSAT HRI pointings  
in the direction of the ``Lockman Hole",
a line of sight with exceptionally low HI column density 
(\cite{Hget98}). 
RXJ1053.7+5735 is
one of the most distant clusters ever selected by diffuse X-ray emission
(for another most distant X-ray selected cluster at z=1.26, 
RXJ0848.9+4452, see Rosati et al. 1999; Stanford et al. 2001). 
Together with the fact that its X-ray morphology is clearly double-lobed and
highly anisotropic, which may well be a sign of a cluster in the making,
the RXJ1053.7+5735 may provide us with unique information to
better understand the evolution of
clusters and galaxies in the early universe.
Throughout the paper, we use $H_{o}$ = 100 $h$ km s$^{-1}$ Mpc$^{-1}$
and $q_{o}$ = 0.5. 

\section{RXJ1053.7+5735}
The X-ray source RXJ1053.7+5735 was discovered 
during our
deep pointings of
1.31 Msec with ROSAT HRI
in the direction of the Lockman Hole.
Figure 1 shows the HRI contours superposed on 
the image composed of V, R, \& I band exposures,
where North is up and East is left.
The HRI image revealed a clearly extended
X-ray emission with unusual double-lobed X-ray morphology.
The angular size of the source is 1.7 $\times$ 0.7 arcmin$^2$
and its X-ray flux is 2 $\times$ 10$^{-14}$ erg cm$^{-2}$ s$^{-1}$
(\cite{Hget98b}).
Our subsequent deep optical/NIR imaging follow-ups (V $<$ 26.5, R $<$ 25, I $<$ 25,
K $<$ 20.5) with LRIS and NIRC on Keck,
and the Calar Alto Omega Prime camera
revealed a bright 7 arcsecond-long arc with an
integrated magnitude of R=21.4, 
and an overdensity of galaxies in both X-ray lobes
(e.g. \cite{Thet01}).
Further Keck LRIS/NIRSPEC spectroscopic observations on the bright arc
and one of the brightest possible member galaxies confirmed that the bright
arc is a lensed galaxy at a redshift z = 2.57
and the galaxy is at a redshift of z = 1.263 (\cite{Thet01}).
Deep VRIzJHK photometry data 
also produced concordant
photometric redshifts for more than 10 objects
at redshift of z $\sim$ 1.3, confirming
that at least the eastern lobe is a massive cluster at high redshift.
The improbability of chance alignment and similarity of colors for the galaxies in the
two X-ray lobes are consistent with the western lobe also being at z $\sim$ 1.3
(see \cite{Thet01} for further details).
The present data are
all consistent with
the identification of the X-ray source with one or two 
high redshift cluster(s) of galaxies.
The spatial distribution of the X-ray gas and cluster galaxies
in this system suggest that it could be a valuable example of
a dynamically unsettled high redshift cluster, 
however, further optical, NIR, \& X-ray investigations are planned 
for a more definitive statement about the dynamical state of this
interesting cluster. 
\begin{figure}
 \resizebox{\hsize}{!}{\includegraphics{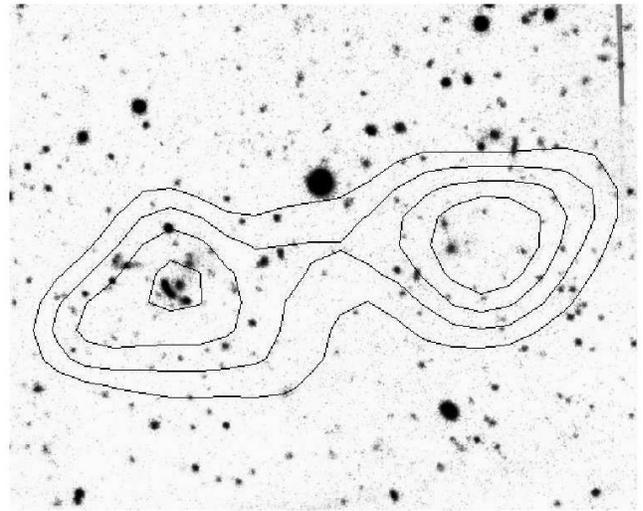}}
 \caption{
The ROSAT HRI contours of the cluster RXJ1053.7+5735
overlaid on 
the image composed of V, R, \& I band images.
North is up and East is left.
}
\label{FigTemp}
\end{figure}

\section{XMM Observations and Data Reduction}
The XMM observations of the cluster RXJ1053.7+5735 were
performed with the European Photon
Imaging Camera (EPIC) during the period April 27- May 19, 2000 
as a part of the performance verification (PV) phase Lockman Hole 
observation centered on the
sky position RA 10:52:43, DEC +57:28:48 (J2000),
for a total exposure time of $\sim$ 190 ksec.   
The cluster center is at an off-axis angle of $\sim$ 10\arcmin\ from the
pointing center, and this causes a vignetting effect, on the average,
of a factor of about two with respect to the pointing center. 
The EPIC cameras were operated in the standard full-frame mode.
The thin filter was used for the pn camera, while the thin and thick
filters were used for the two MOS cameras.  
The pn detector consists of 12 CCDs each 13.6 $\times$ 4.4 arcmin,
with energy resolutions of 6.7\% at
1 keV, while MOS detectors consist
of 7 CCDs each 10.9 $\times$ 10.9 arcmin  
with energy resolutions of 5.7\% at 1 keV.
The sensitive FOV of both detector is about 30\arcmin diameter and
the energy range is both 0.1 to 15 keV.
(For further details of the PV Lockman Hole observation, see
\cite{Hget01})

Light curves were visually inspected,
and time intervals with high background (0.5-10 keV count rate higher than 
8 cts/s for the pn and 3 cts/s for MOS) were excluded. 
We used both single and double events.
The remaining effective exposure times for each of three detectors were
approximately 100 ksec.
The astrometries calculated from the WCS keywords were offset by 
5\arcsec-25\arcsec\ from the known optical counterparts. Thus, a transversal
shift
and
a rotation angle were fit for each dataset, leading to residual systematic
position errors of 1\arcsec-2\arcsec.

Fig. 2 shows the exposure corrected XMM image of RXJ1053.7+5735 
in the 0.5-2 keV band, an energy range where the bulk of the redshifted
cluster emission
falls upon and therefore the contrast against the background is the strongest. 
The image was created by combining all events from three
(pn, MOS1, \& MOS2) cameras.  
The raw data were smoothed with a Gaussian with $\sigma$ = 7\arcsec. 
The lowest contour is 2$\sigma$ above the background, and
the contour interval is 0.5$\sigma$.
A combined exposure map was calculated for the pn plus MOS cameras.

Fig. 2 clearly shows that the cluster emission is extended and double-lobed,
consistent with the ROSAT HRI image, and
its shape is indicative of non-regularity of this cluster.
Unfortunately, the pointing variation between the different 
revolutions was not large enough to clearly remove the effect of 
inter-CCD gaps, 
which happen to run through the cluster, particularly around the eastern lobe,
and this somewhat distorted the
level and shape of the eastern-lobe contours.

\begin{figure}
 \resizebox{\hsize}{!}{\includegraphics{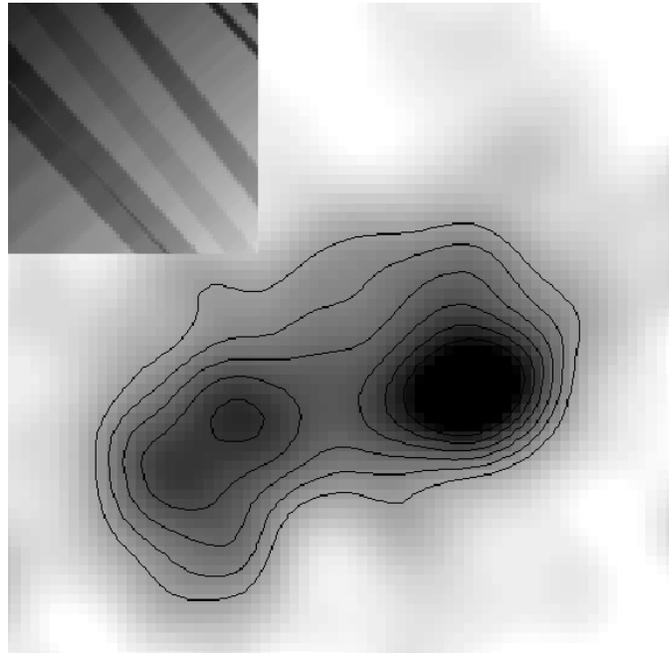}}
 \caption{ 
The exposure corrected XMM image of the cluster RXJ1053.7+5735
in the 0.5-2 keV band.
The image was created by combining all events from three
(pn, MOS1, \& MOS2) cameras.
North is up and East is left.
The raw data were smoothed with a Gaussian with $\sigma$ = 7\arcsec.
The lowest contour is 2$\sigma$ above the background, and
the contour interval is 0.5$\sigma$.
Inset shows a combined (pn+MOS) exposure map (0.5-2 keV) for the 
area identical to the cluster image.
Inter-CCD gaps happen to run through the cluster, in particular through 
the eastern lobe, and this
somewhat distorted the level and shape of the contours.
}
\label{FigTemp}
\end{figure}

\section{Spectral Analysis}

The spectra were extracted from an elliptical region centered at the
cluster image with a semi-major/minor axis 0\farcm9 \& 0\farcm6, respectively
(corresponding to 0.23 \& 0.16 Mpc/h at z= 1.263). 
The backgrounds were estimated from a region of the detector 
at the same off-axis angle
surrounding the  cluster, after removing point sources.
We obtain  791$\pm$37 and 519$\pm$30 net counts (in the 0.2-10.0 keV band)
for pn and MOS(1+2), respectively. 
 We regrouped the counts in order to have a S/N $\geq$ 3 in each bin after the
background subtraction.
We fitted the spectra with a
thermal plasma emission model from Raymond \& Smith(1977) using the 
XSPEC 11.0 with respective response matrices (Haberl 2001). 
For the MOS cameras, two filters were used alternately, thus  we summed 
all the MOS spectra according to  the used filter and fit two sets of spectra 
jointly.
We fixed a characteristic column density of neutral
hydrogen $N_{\rm {H}}$ at a value of 5.6 $\times$ 10$^{19}$ cm$^{-2}$ and redshift at z = 1.263.
For the case of a joint fit, 
each spectrum was fitted with its own normalization, but 
with common 
temperature (and metallicity, if applicable).

\begin{figure}[h]
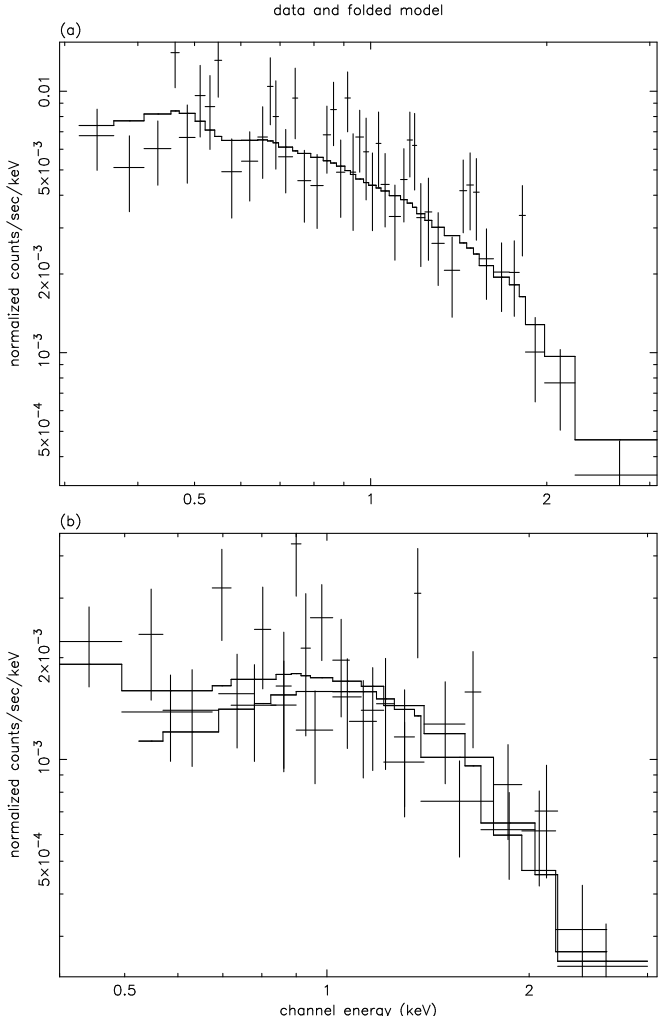

\resizebox{\hsize}{!}{\includegraphics[angle=-90,bb=0 30 555 700,clip]{s229pn.lda.ps}}
\resizebox{\hsize}{!}{\includegraphics[angle=-90,bb=60 30 570 700,clip]{s229m_thin+s229m_thic.lda.ps}}
\caption{ Rebinned spectra and best-fit models for cluster RXJ1053.7+5735
 with pn (a) and MOS1+2 (b) cameras.
 The crosses are the observed spectrum and the solid line denotes
 best-fit model.
}
\label{FigTemp}
\end{figure}
\begin{figure}[h]
 \resizebox{\hsize}{!}{\includegraphics[angle=-90,bb=81 30 570 700,clip]{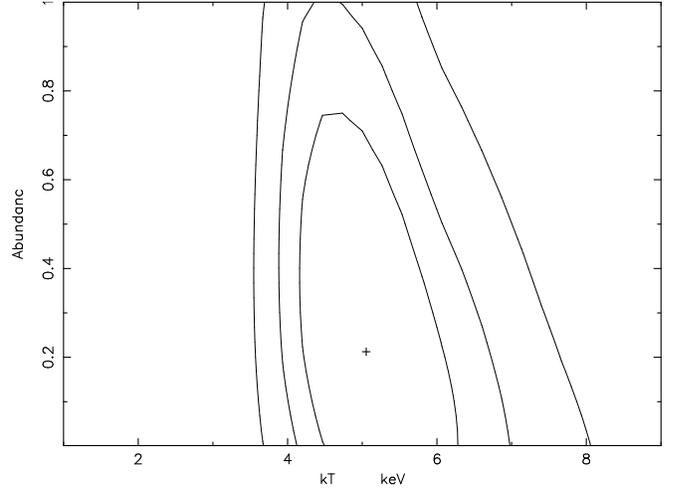}}
 \caption{
Two-dimensional $\chi^{2}$ contours at 68.3\%, 90\%, and 99\%
confidence levels ($\Delta\chi^{2}$=2.30, 4.61 and 9.21) for the
cluster RXJ1053.7+5735 temperature and abundance
(in solar units) based on the joint fit for  pn and MOS all together.
}
\label{FigTemp2}
\end{figure}
\begin{figure}[h]
 \resizebox{\hsize}{!}{\includegraphics[angle=90,bb=30 30 580 700,clip]{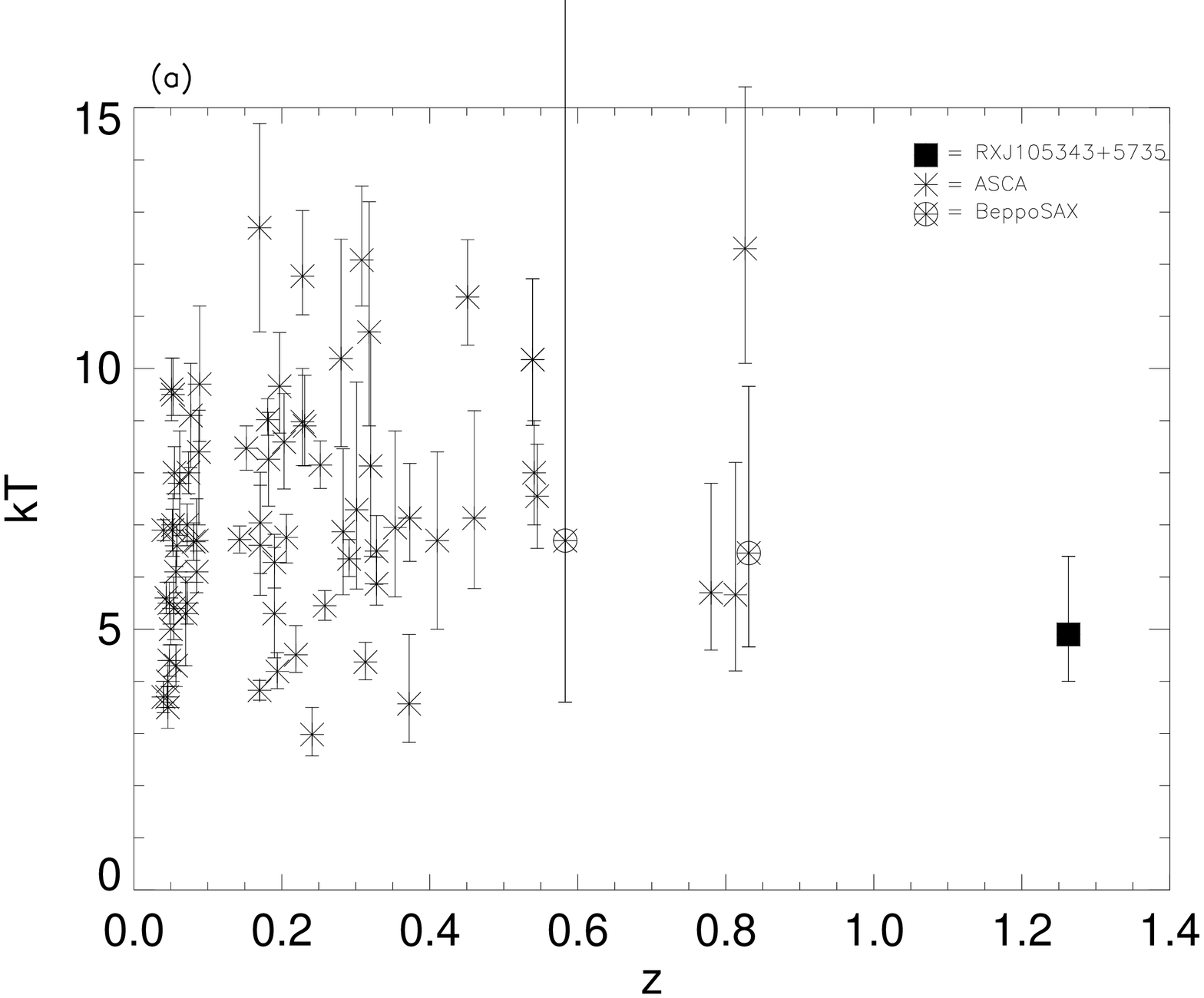}}
 \resizebox{\hsize}{!}{\includegraphics[angle=90,bb=30 30 580 700,clip]{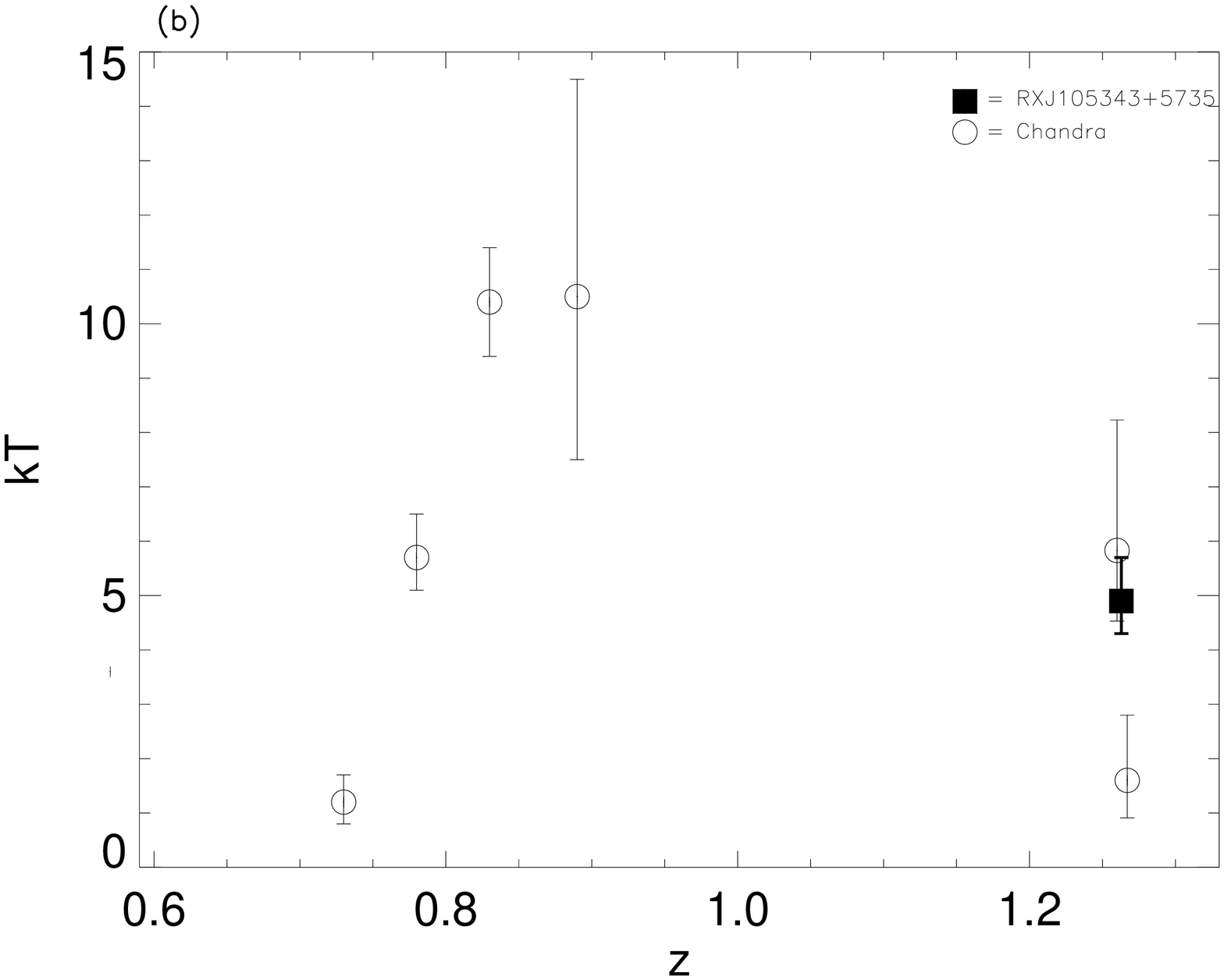}}
 \caption{
(a): 
New XMM temperature and its 2$\sigma$ error (90\%) for RXJ1053.7+5735 plotted
with other clusters observed by past satellites over various redshifts,
taken from the literature 
(\cite{Ma98}; Mushotzky \& Scharf 1997; Donahue et al. 1999; Della Ceca et al.
2000).
Note that the temperature error for XMM measurement
is comparable to those by ASCA for clusters at z $\sim$ 0.4.
Moreover, the measured temperature and luminosity showed that
one can easily reach the intrinsically X-ray
faint and cool cluster regime
comparable with those of z $\sim$ 0.4 clusters observed by past satellites.
(b):
Comparison between XMM temperature (RXJ1053.7+5735) and new 
Chandra measurements 
of other high-z clusters (z $>$ 0.6) from Borgani et al. (2001). 
The error of RXJ1053.7+5735 is changed to 1$\sigma$ (68.3\%)
to match quoted errors in Borgani et al. (2001).
}
\label{FigTemp}
\end{figure}

\begin{table}[h]
\caption[]{Results of Spectral Fits for Cluster RXJ1053.7+5735}
\begin{tabular}{lrrlrr}
\hline
\hline
\noalign{\smallskip}
Detector      & kT$^a$ & Abundance   &  $\chi^{2}$(dof) \\
              & (keV) & ($\odot$) &  &                     \\
\noalign{\smallskip}
\hline
\noalign{\smallskip}
pn      & 4.8 (3.7-6.3) &  0.3(fixed)   & 0.86(43) \\
MOS1+2  & 4.9 (3.6-7.9) &  0.3(fixed)   & 0.78(29) \\
ALL     & 4.9 (4.0-6.4) &  $<$0.62       & 0.83(72) \\
\noalign{\smallskip}
\hline
\end{tabular}

$^a$ The quoted temperature errors correspond to a single parameter error at 90\% confidence \\
\label{tab:det}
\end{table}

\begin{table}[h]
\caption[]{Comparison Between the Two Lobes (pn+MOS1+MOS2)}
\begin{tabular}{lrrlrr}
\hline
\hline
\noalign{\smallskip}
Lobe         & kT  & Lbol         &Abundance   &  $\chi^{2}$(dof) \\
              & (keV) & (10$^{44}$erg/s) & ($\odot$)   &                       \\
\noalign{\smallskip}
\hline
\noalign{\smallskip}
East  & 3.7 (2.7-4.9) &  1.5 & 0.3(fixed)       & 1.03(30) \\
West  & 6.2 (4.4-10.4) & 1.9 & 0.3(fixed)       & 0.82(43) \\
E+W   & 4.9 (4.0-6.4) &  3.4 & $<$0.62          & 0.83(72) \\
\noalign{\smallskip}
\hline
\end{tabular}
\label{tab:det}
\end{table}

In Fig 3, and Table 1, the results of the best-fits are shown.
Fig. 4 shows the two-parameter $\chi^{2}$ contours for the cluster metallicity 
and Tx.
The best fit temperature based on a simultaneous fit of all (pn+MOS)
spectra is 
4.9 $^{+1.5}_{-0.9}$ keV.
All temperature uncertainties quoted are at the 90\% confidence levels for a 
one-dimensional fit ($\Delta\chi^{2}$=2.71).
No prominent iron K$\alpha$ complex is visible at $\sim$ 3 keV 
(redshifted 6.7 keV). At kT$\sim$ 5 keV, and the cluster redshift, only
the K$\alpha$ is strong enough to constrain the abundance 
with our spectrum.
As a result, metallicity is poorly constrained even using the joint fit of
all spectra, 
with an upper limit on the single iron abundance of 0.62 solar 
(with 68\% confidence). 
Despite the fact that the cluster was observed at fairly large off-axis angle,
the temperature errors are much smaller compared with 
those of typical
measurements based on ASCA or Beppo-Sax observations of 
high-z (z $>$ 0.6) clusters (Fig. 5a), 
and the errors 
are comparable with typically ``on-axis'' Chandra 
observations of z $\sim$ 0.8  clusters (Fig. 5b),
demonstrating the power of XMM for
determining the X-ray temperature for high-z clusters. 
Moreover, the measured temperature and luminosity show that
one can easily reach the intrinsically X-ray 
faint and cool cluster 
regime comparable with those of z $\sim$ 0.4 clusters observed by 
past satellites, which enables the investigation of the evolution
of various cluster X-ray properties 
without additional
evolutionary assumptions.
Vignetting correction is estimated by using the average vignetting model
between 0.5-7 keV, also averaged over the spectral extraction region.
Using the best fit model parameters, we derived an 
unabsorbed (0.2-10) keV 
flux of $f_{0.2-10}$ = 
3.0 $\times$ 
10$^{-14}$ erg cm $^{-2}$ s$^{-1}$,
corresponding to a
luminosity in the cluster rest frame of 
$L_{0.2-10}$ = 
3.1$\times$ 
10$^{44}$ h$_{50}^{-2}$erg s$^{-1}$ and a bolometric luminosity of
$L_{bol}$ = 
3.4 $\times$ 
10$^{44}$ h$_{50}^{-2}$erg s$^{-1}$.
Due to the still preliminary status of the
current EPIC calibration, and some simplification about the vignetting
correction, we estimate systematic flux errors
on the order of 10\%. 

Temperature and luminosity are also estimated for each (eastern and  western) 
lobe of RXJ1053.7+5735, using a circular (radius = 0\farcm47) extracting region centered at each lobe.
Our fitting for each lobe using all data (pn+MOS) shows that Tx is
3.7 $^{+1.2}_{-1.0}$ keV and 6.2 $^{+4.2}_{-1.8}$ keV, while
L$_{bol}$ is 1.5 $\times$
10$^{44}$ h$_{50}^{-2}$erg s$^{-1}$
and 1.9
$\times$
10$^{44}$ h$_{50}^{-2}$erg s$^{-1}$, for eastern and western
lobes, respectively (Table 2). 
For any other regions, such as the bridge between the two lobes,
the counts are less than 10\% of total cluster counts and not sufficient
for any reasonable Tx investigations.

\section{Lx-Tx relation}

\begin{figure}[h]
 \resizebox{\hsize}{!}{\includegraphics[angle=90,bb=50 200 550 750,clip]{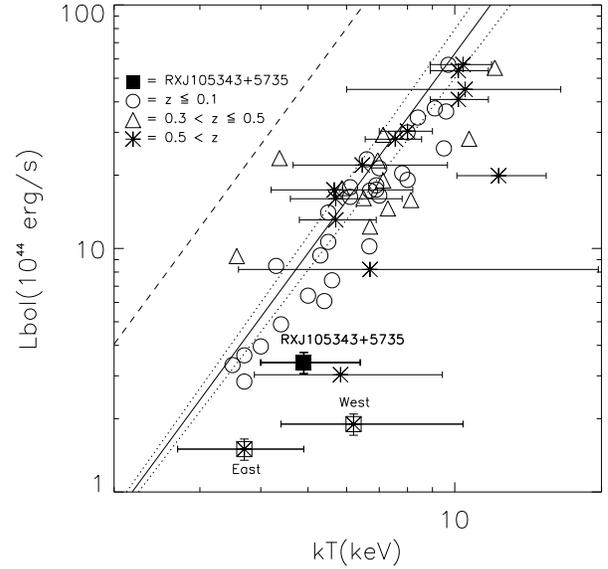}}
 \caption{
Newly measured X-ray properties for a high-z (z = 1.263)
cluster, RXJ1053.7+5735
(the square symbol),
is compared
with other clusters at various redshifts.
The error on the temperature represents the 90\% confidence range.
The error on  the RXJ1053.7+5735 luminosity is 10\%.
The eastern and western lobes of RXJ1053.7+5735 (denoted as ``East'' and ``West'')
are also shown.
The solid line is the L$_{bol}$-Tx relation of
 L$_{bol}$ = 10$^{-0.92}T^{2.72} $
from Wu et al. (1999) together with $\pm$ 2$\sigma$ lines (dotted lines).
The dashed line represents the evolving L$_{bol}$-Tx relation
in the form of
L$_{bol}$ $\propto$ (1+z)$^{2}$ at z = 1.263, which required to
make the observed non-evolving XLF to be consistent with a $\Omega_{m}$ = 1
universe.
 The circles, triangle, \& stars denote low redshift (z $\leq$ 0.1),
 intermediate redshift (0.3 $<$ z $\leq$ 0.5), and high redshift
(0.5 $<$  z) clusters, respectively.
For brevity, temperature error bars of only high-z clusters are plotted.
The new cluster temperature and L$_{bol}$ we have measured for RXJ1053.7+5735
is consistent with a weak/no  evolution of the L$_{bol}$ - Tx
relation out to  z $\sim$ 1.3, which lends support to a
low $\Omega_{m}$ universe,
although more data-points of z $>$ 1 clusters are required for a more definitive statement.
The caution has to be also exercised
in interpreting the result, because of the uncertainty associated with
the dynamical status of this cluster.
}
\label{FigTemp}
\end{figure}

 Clusters exhibit a correlation between their X-ray
luminosity (Lx) and temperature (Tx). This correlation is often
used to compare  gas and dark matter contents in clusters, as Lx 
is 
connected with the gas while Tx is related
to the total gravitating mass of the cluster.
The cluster Lx-Tx relationship therefore,
can be used to examine the evolution of clusters and should be 
able to tell the range of redshifts where the cluster formation 
has taken place (e.g. Scharf \& Mushotzky 1997).

At low redshift (z $<$ 0.1), the Lx-Tx relation is well measured
(e.g. Mushotzky 1984; Edge \& Stewart 1991; David et al. 1993)
and expressed via power law: Lx $\propto$ Tx$^{\alpha}$, with 
$\alpha \sim $ 2.6-2.9 (\cite{Ma98}; Allen \& Fabian 1998;
Jones \& Forman 1999; Arnaud \& Evrard 1999).

At higher redshifts, however, 
no significant difference from the low redshift relations are
reported, despite the fact that
a relatively large evolution of the Lx-Tx
relation is required in order to account,
within the frame work of the critical universe, 
for the weak
evolution observed in the cluster X-ray luminosity function (XLF) 
out to z $\sim$ 0.8 (Rosati et al. 1998). 
Mushotzky \& Scharf (1997) (see also Henry 2000), 
by comparing the clusters at z $<$ 0.1 with those at z $>$ 0.1,
found no convincing evidence for a significant evolution 
of Lx-Tx relationship out to z $\sim$0.4-0.5,
which is consistent with recent result of
Wu et al. (1999) where they used the 
largest sample of clusters to date from the literature, and estimated
$\alpha \sim $ 2.7.
For even higher redshifts,
Donahue et al. (1999), 
using a complete sample of (five) high redshift (z $>$ 0.5) EMSS clusters,
have detected no significant evolution up to  z $\sim$ 0.8 
(see also Della Ceca et al. 2000; Borgani et al. 2001).

 In Fig. 6, we compare our newly measured X-ray properties for
RXJ1053.7+5735 with other clusters at 
various redshifts from the literature 
(\cite{Ma98}; Mushotzky \& Scharf 1997; Donahue et al. 1999; Della Ceca et al.
2000; Borgani et al. 2001). 
The square denotes the new L$_{bol}$ and Tx for RXJ1053.7+5735.
The error on the temperature represents the one dimensional 
90\% confidence range.
The error bar on the RXJ1053.7+5735 luminosity is 10\%.
The eastern and western lobes of RXJ1053.7+5735 (denoted as ``East'' and ``West'')
are also separately shown.
The solid line is the L$_{bol}$-Tx relation of
 L$_{bol}$ = 10$^{-0.92}T^{2.72}$ 
from Wu et al. (1999) together with $\pm$ 2$\sigma$ lines (dotted lines).
The dashed line represents the evolving L$_{bol}$-Tx relation
in the form of
L$_{bol}$ $\propto$ (1+z)$^{2}$ at z = 1.263, which would be required to
make the observed weak-evolving XLF to be consistent with a $\Omega_{m}$ = 1
universe 
(\cite{Boet99}).
 The circles, triangle, \& stars denote low redshift (z $\leq$ 0.1),
 intermediate redshift (0.3 $<$ z $\leq$ 0.5), and high redshift 
(z $>$ 0.5) clusters, respectively. 
Total cluster luminosity may be somewhat underestimated because of
the fundamental difficulty in measuring the
luminosity of an extended source using a finite-sized  extraction region.
However, 
the treatments (if any)
against the missing flux outside the extraction region
often require difficult extrapolation and are not uniform 
for the low-z cluster
samples, or other high-z samples in the literature.
We therefore chose not to include the correction associated with the
missing flux.
Nevertheless, we crudely estimate the effect in our cluster
using a profile from Neumann \& Arnaud (1999). We find that
the uncertainty in Lx is on the order of 10\%, which is
similar with the size of error bar shown in Fig. 6.
The new cluster temperature and L$_{bol}$ we have measured for RXJ1053.7+5735
is in agreement with other high-z Lx-Tx analyses (e.g. Donahue et al 1999; Della Ceca et al. 2000; Borgani et al. 2001).

\section{Discussion}

The dynamical state of high-z clusters provides us with  valuable and
direct information on the formation and evolution of clusters.  In
hierarchical models, structures form from the bottom-up and thus, for
many clusters, ``formation'' means an ongoing sequence of mergers and
interactions with other clusters and groups.  
X-ray signatures of cluster interaction include 
surface brightness anomalies (e.g. non-symmetric, non-isotropic
X-ray isophotes; substructures; elongate X-ray core;
 an X-ray peak offset from
the peak of the galaxy distribution), and 
non-isothermal \& asymmetric temperature distributions. Evidence for
such cluster substructures and merger events in the local universe
have been obtained  with ROSAT and ASCA satellite (see Buote 2001 for
a review).
For higher redshifts (z $>$ 0.5), the majority of clusters show clearly 
distorted X-ray
morphology (e.g. Neumann \& B\"ohringer 1997; Gioia et al. 1999; Della Ceca
et al. 2000; Neumann \& Arnaud 2000; Stanford et al. 2001; Jeltema et
al. 2001), suggesting that they are in an unrelaxed state.

        The fact that the X-ray morphology of RXJ1053.7+5735 is
double-lobed suggests that we may be seeing a merger event at
z $\sim$ 1.3, although our optical/NIR data are currently insufficient for
a definitive statement about the dynamical state of this cluster. 
As for the Tx distribution, 
Fig.7 shows the contour plot of an image in the 1-2 keV band overlaid on the
gray-scale image in the 0.5-1 keV band. The figure shows some hint of
a difference in the spatial distribution
of the 1-2 keV image contour with respect to the 0.5-1 keV map,
which may be interpreted as a Tx variation.
The western lobe also shows a hint of higher Tx, however, it
is statistically inconclusive (Table 2). 
Future XMM (or Chandra) data of this cluster are
needed, as well as more optical/NIR data, 
to improve our knowledge of the dynamical state of this cluster.

\begin{figure}
 \resizebox{\hsize}{!}{\includegraphics{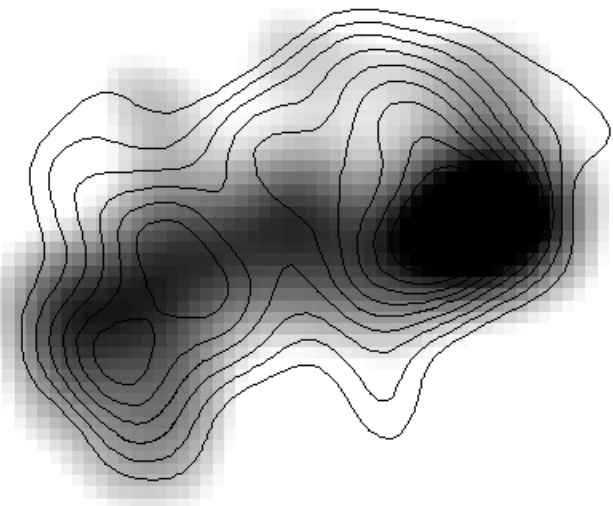}}
 \caption{
The contour plot of 1-2 keV image overlaid on the
gray-scale 0.5-1 keV image.
Both images are smoothed  with a Gaussian with $\sigma$ = 7\arcsec.
}
\label{FigTemp}
\end{figure}

The new cluster temperature and Lx we have measured for
RXJ1053.7+5735 is consistent with a weak/no evolution of the Lx-Tx
relation out to z $\sim$ 1.3, whereas a strong evolution of the Lx-Tx
relation is required to make the observed weak-evolving XLF to be
consistent with a $\Omega_{m}$ = 1 universe. Thus, our results could be
interpreted as a support for a low $\Omega_{m}$ universe, although 
more data-points of z $>$ 1 clusters are needed for a more definitive 
statement.
The caution has to be also exercised in interpreting the result,
because of the uncertainty associated with the dynamical state of the
cluster. In fact, our results may emphasize the fundamental difficulty in
constraining the cosmology from the evolution of the XLF and the Lx-Tx
relation, since the baseline of the comparison, the local Lx-Tx
relation and the theoretical M-T relationships, are essentially valid
for dynamically relaxed clusters (see also Ricker \& Sarazin 2001).
If the majority of high-z clusters are in the unrelaxed state,
the impact of dynamically unrelaxed cluster states has to be carefully assessed. 
In particular, during
mergers, clusters are expected to follow a complex track in the Lx-Tx
plane as shown in the numerical simulations (Ricker \& Sarazin 2001),
and therefore, 
the relatively low luminosity of
RXJ1053.7+5735 in view of its temperature may be interpreted
as further evidence for the unrelaxed state of the cluster.

\begin{acknowledgements}
We thank Pat Henry for a careful reading of the paper and useful comments.
We also thank S. Majerowicz for his help in analyzing the MOS spectra.
We acknowledge referee's comments which improved the manuscript.
Part of the work was supported by the
German {Deutsches Zentrum f\"ur Luft- und Raumfahrt} DLR project numbers
 50 OX 9801 and 50 OR 9908.
\end{acknowledgements}


\begin{thebibliography}{}

\bibitem[Allen \& Fabian 1998]{AlFa98}
Allen S.W., \& Fabian A.C. 1998, MNRAS, 297, L57

\bibitem[Arnaud \& Evrard 1999]{ArEv99}
Arnaud M., \& Evrard A.E. 1999, MNRAS, 305, 631

\bibitem[Borgani et al. 1999]{Boet99}
Borgani S., Rosati P., Tozzi P., Norman C., 1999, ApJ 517, 40

\bibitem[Borgani et al. 2001]{Boet01}
Borgani S., Rosati P., Tozzi P., Stanford S.A. et al. 2001, (astro-ph/0106428)


\bibitem[Buote 2001]{Bu01}
Buote D.A. 2001, To appear in Merging Processes in Clusters of Galaxies,
ed. L. Feretti, I.M. Gioia \& G. Giovannini (Dordrecht: Kluwer)

\bibitem[Burke et al. 1997]{Buet97}
Burke D.J., Collins C.A., Sharples R.M., Romer A.K., 
   Holden B.P., Nichol R.C., 1997, ApJ 488, L83

\bibitem[Cagnoni et al. 2001]{Caet01}
Cagnoni I., Elvis M., Kim D.-W., Mazzotta P. et al. 2001, ApJ in press (astro-ph/0106066)

\bibitem[David et al. 1993]{Daet93}
David L.P., Slyz A., Jones C., Forman W. Vrtilek S.D. \& Arnaud K.A. 1993, ApJ 412, 479

\bibitem[Della Ceca et al. 2000]{Deet00}
Della Ceca R., Scaramella R., Gioia I.M., Rosati P., Fiore F., \& Squires G.
2000 A\&A 353, 498

\bibitem[Donahue et al. 1999]{Doet99}
Donahue M., Voit G.M., Scharf C.A.,Gioia I.M., Mullis C.R., Hughes J.P., \&
Stocke J.T. 1999, ApJ 527, 525  

\bibitem[Edge \& Stewart 1991]{EdSt91}
Edge A.C., \& Stewart G.C. 1991, MNRAS, 252, 428

\bibitem[Edge, Stewart, \& Fabian 1992]{EdStFa92}
Edge A.C., Stewart G.C., \& Fabian A.C., 1992, MNRAS, 255, 431

\bibitem[Eke et al. 1996]{Ekeet96}
Eke V.R., Cole S., Frenk C.S., 1996, MNRAS 282, 263


\bibitem[Gioia et al. 1999]{Giet99}
Gioia, I.M., Henry, J.P., Mullis, C.R., Ebeling, H., \& Wolter, A., 1999 ApJ, 117,
2608

\bibitem[Haberl 2001]{Ha01}
Haberl F., 2001 private communication

\bibitem[Hasinger et al. 1998a]{Hget98}
Hasinger G., Burg R., Giacconi R., et al., 1998a, A\&A 329, 482 

\bibitem[Hasinger et al. 1998b]{Hget98b}
Hasinger G., Giacconi R., Gunn J.E., et al., 1998b, A\&A 340, L27

\bibitem[Hasinger et al. 2001]{Hget01}
Hasinger G., Altieri B., Arnaud M. et al., 2001, A\&A 365, L45 


\bibitem[Henry 2000]{He00}
Henry J.P.,2000, ApJ 534, 565

\bibitem[Jeltema et al.]{Jeet01}
Jeltema T.E., Canizares C.R., Bautz M.W., Malm M.R., Donahue M., Garmire G.P.
2001, ApJ in press

\bibitem[Jones \& Forman 1999]{JoFo99}
Jones C., Forman W., 1999, ApJ 511, 65


\bibitem[Markevitch 1998]{Ma98}
Markevitch, M. 1998, ApJ 504, 27 


\bibitem[Mushotzky 1984]{Mu84}
Mushotzky R., 1984, Phys. Scr. T7, 157


\bibitem[Mushotzky \& Scharf 1997]{MuSc97}
Mushotzky R., Scharf C., 1997, ApJ 482, L13

\bibitem[Neumann \& Bohringer 1997]{NeBo99}
Neumann, D.M., B\"ohringer, H., 1997, MNRAS 289, 123

\bibitem[Neumann \& Arnaud 1999]{NeAr99}
Neumann D.M., Arnaud M., 1999, A\&A 348, 711 

\bibitem[Neumann \& Arnaud 2000]{NeAr00}
Neumann D.M., Arnaud M., 2000, ApJ 542, 35 

\bibitem[Peebles et al. 1989]{Peet89}
Peebles P.J.E., Daly R., Juskeiewiez R., 1989, ApJ 347, 563 

\bibitem[Press \& Schechter 1974]{PrSc74}
Press W., Schechter P., 1974, ApJ 187, 425

\bibitem[Raymond \& Smith 1977]{RaSm77}
Raymond J.C., Smith B.W., 1977, ApJS 35, 419

\bibitem[Ricker \& Sarazin 2001]{RiSa01}
Ricker P.M., Sarazin C.L., 2001 ApJ in press

\bibitem[Rosati et al. 1995]{Roet95}
Rosati P., Della Ceca R., Burg R., Norman C., Giacconi R., 1995, ApJ 445, L11

\bibitem[Rosati et al. 1998]{Roet98}
Rosati P., Della Ceca R., Norman C., Giacconi R., 1998, ApJ 492, L21

\bibitem[Rosati et al. 1999]{Roet99}
Rosati P., Stanford S.A., Eisenhardt P.R., Elston R. et al., 1999, AJ 118, 76

\bibitem[Scharf \& Mushotzky 1997]{ScMu97}
Scharf C.A., Mushotzky R. 1997, ApJ 485, L65

\bibitem[Scharf et al. 1998]{Scet98}
Scharf C.A., Jones L.R., Ebeling H., et al., 1997, ApJ 477, 79

\bibitem[Stanford et al. 2001]{Stet98}
Stanford S.A, Holden B., Rosati P., et al. 2001, ApJ, in press, astro-ph/0012250 


\bibitem[Thompson et al. 2001]{Thet01} 
Thompson D., Pozzetti L., Hasinger G, et al., 2001, A\&A (submitted)

\bibitem[Vikhlinin et al. 1998]{Viet98} 
Vikhlinin A., McNamara B.R., Forman W., et al., 1998, ApJ 502, 598


\bibitem[Wu et al. 1999]{Wuet99} 
Wu X-P., Xue Y-J., Fang L-Z., 1999, ApJ 524, 22



\newpage 

\end{thebibliography}
\end{document}